\documentstyle[12pt,epsf,epsfig]{article}
\def\ga{\mathrel{\raise.3ex\hbox{$>$\kern-.75em\lower1ex\hbox{$\sim$}}}}
\def\la{\mathrel{\raise.3ex\hbox{$<$\kern-.75em\lower1ex\hbox{$\sim$}}}}
\def\gyr{{\rm \, G\kern-0.125em yr}}
\def\gev{{\rm \, Ge\kern-0.125em V}}
\def\tev{{\rm \, Te\kern-0.125em V}}
\def\beq{\begin{equation}}
\def\eeq{\end{equation}}
\def\ss{\scriptscriptstyle}

\def\mchi{m_{\tilde \chi}}

\def\m12{m_{1\!/2}}

\def\mfb{\overline{m}_{\ss f}}
\def\mf{m_{\ss{f}}}

\def\gf{\gamma_f}
\def\thm{\theta_\mu}
\def\tha{\theta_A}
\def\thb{\theta_B}

\def\cp{C\!P}
\def\ohsq{\Omega_{\widetilde\chi}\, h^2}

\begin{document}
\begin{titlepage}
\pagestyle{empty}
\baselineskip=21pt
\rightline{UMN--TH--1705/98}
\rightline{MADPH-98-1059}
\rightline{June 1998}
\vskip1.25in
\begin{center}

{\large{\bf More on Electric Dipole Moment Constraints
on Phases in the Constrained MSSM}}
\end{center}
\begin{center}
\vskip 0.5in
{Toby Falk}

{\it Department of Physics, University of Wisconsin, Madison, WI~53706,
WI, USA}

 and 

Keith A.~Olive

{\it
{School of Physics and Astronomy,
University of Minnesota, Minneapolis, MN 55455, USA}\\}
\vskip 0.5in
{\bf Abstract}
\end{center}
\baselineskip=18pt \noindent
We reconsider constraints on $\cp$-violating phases in
mSUGRA, sometimes also referred to as the
Constrained Minimal Supersymmetric Standard Model. We include the recent 
calculations of Ibrahim and Nath on the chromoelectric and purely gluonic
contributions to the quark electric dipole moment and combine cosmological
limits on gaugino masses with experimental bounds on the neutron (and electron)
electric dipole moments. The constraint on the phase of the Higgs mixing mass
$\mu$,
$|\thm|$, is dependent on the
value of the trilinear mass parameter, $A$, in the model and on
$\tan \beta$. For values of $|A| < 300 \gev$ at the GUT scale, we find
 $|\thm|/\pi \la 0.05$, while for $|A| < 1500 \gev$, $|\thm|/\pi \la 0.3$.
Thus, we find that in principle, large CP violating phases are compatible with
the bounds on the electric dipole moments of the neutron and electron, as well
as remaining compatible with the cosmological upper bound on the relic density
of neutralinos.  The other
$\cp$-violating phase $\tha$ is essentially unconstrained.
\end{titlepage}
\baselineskip=18pt
There has been considerable progress recently in establishing strong
constraints on the supersymmetric parameter space from recent runs at LEP
\cite{efos2,efgos}. These constraints provide a lower bound to the neutralino
mass of $\sim 40\gev$, when in addition to the bounds from experimental
searches for charginos, associated neutralino production and Higgs bosons, 
constraints coming from cosmology and theoretical simplifications concerning
the input scalar masses in the theory are invoked.   Indeed, when all
soft supersymmetry breaking scalar masses are set equal to each other at the
GUT scale (including the soft Higgs masses), an ans\"{a}tz
typically referred to as mSUGRA,
this full set of constraints excludes
 the low values of
$\tan \beta < 2$ for $\mu < 0 $ and $< 1.65$ for $\mu > 0$. 
MSUGRA is interesting because it is simple, predictive and naturally
provides a stable dark matter candidate
\cite{susygut} (an LSP bino-type neutralino) over most of its parameter space.

MSUGRA (as well as the MSSM, which does not necessarily make any a priori
assumption concerning the scalar masses other than the constraint that they should
not lead to charge and color breaking minima \cite{us,someothers}) is
well known to contain several independent $\cp$-violating phases. If we assume
that all of the supersymmetry breaking trilinear mass terms, $A_i$ are equal at
the scale $M_X$ at which the mass parameters unify, then
the number of independent phases reduces to 2, which we can take as $\tha$ and
$\thm$.  In principle, the supersymmetry breaking bilinear mass term $B$ can
also have a complex phase.   However, the phase of $\mu$ can
always be adjusted so that $\theta_B = - \thm$ by rotating the Higgs fields so
that their vacuum expectation values are real \cite{dgh}, and so 
there are only two independent physical phases, which we take as above.  It is well known
that these phases can lead to sizable contributions to the neutron and
electron dipole moments \cite{dn}.  
To suppress the electric
dipole moments, either large scalar masses (approaching $1$ TeV) or small
angles (of order $10^{-3}$, when all SUSY masses are of order 100 GeV) are
required. It has been commonplace to assume the latter, though the
possibilities for large phases was considered in \cite{ko} and more recently in
\cite{in2}. To reconcile large phases with small electric dipole moments, some
of the sparticle masses are required to be heavy. In \cite{ko}, either large
sfermion or neutralino masses (or both) were required.

 However, unless
$R$-parity is broken and the LSP is not stable, one requires that the
sfermions be heavier than the neutralinos.   If they are much heavier, this
would result in an excessive relic density of neutralinos.  In
\cite{fkos}, we showed that there was a strong correlation between the
$\cp$-violating phases in the MSSM, the cosmological LSP relic density and the
neutron and electron electric dipole moments, when all of the
parameters are set at the weak scale and hence are independent of any RGE
evolution.  Large CP violating phases can also have an important impact on elastic
scattering cross-sections used to calculate dark matter detection rates
\cite{ffo}.  In \cite{fko1},  we further showed that in mSUGRA,
 by combining cosmological constraints on
the mass of the LSP with experimental bounds on the neutron and
electron EDM's, we can bound one of the two new phases in mSUGRA to
lie within $|\thm|<\pi/10$ for $A_0\equiv |A({\rm M_X})|\la 1\tev$.  
The calculation of the contributions to the neutron and electron electric dipole
moments has recently been reinvestigated in \cite{in2,in1} correcting some
errors in \cite{ko} and including the contributions from the chromoelectric and
purely gluonic operators (for the neutron only) which were previously thought
to be small \cite{ads}. These operators, can in fact lead to cancellations
which enlarge the allowable range for the CP violating phases.  Some analytic 
approximations to the cancellations in the edms were 
done in \cite{Roc}.  Here, we wish
to reexamine this problem in light of the recent calculations of \cite{in2,in1}
and to address some questions regarding the independence of the two phases
when the RGE evolution is taken into account \cite{gw}.

In mSUGRA, once the gaugino, soft scalar masses, $A$ and $B$-terms and phases
are given at $M_X$, they can be RGE evolved to the electroweak scale.  In
practice, we use the one-loop RGEs for the masses and two-loop RGEs for the
gauge and Yukawa couplings \cite{ikkt}. The structure of the equations for the
gauge couplings, gaugino masses and the diagonal elements of the sfermion
masses are such that they are entirely real.  The evolutions of the $A_i$,
however, are more complicated, as the $A_i$ pick up both real and
imaginary contributions.  For example, the evolution of $A_t$ is given
by
\begin{eqnarray}
  {dA_t \over dt} = {1\over{8\pi^2}}\left(-{16\over3}\,g_3^2\, M_3
    -3g_2^2\, M_2-{13\over9}\, g_1^2\, M_1 + h_b^2 A_b + 6 h_t^2 A_t\right)
\end{eqnarray}
As one can see, $A_t$ receives real contributions $c_i M$ proportional
to the gaugino mass (whose coefficients $c_i$ are different for each
sfermion in a generation) and (in principle complex) contributions
$d_i h_f^2 A_f$ from the heavy generation (whose coefficients $d_i$
differ for the first two and the third generations); the phases (and
magnitudes) of the $A_i$ must therefore be run separately.  At one
loop, the evolution equation for $\mu$ is given by
\begin{eqnarray}
  {d\mu \over dt} = {\mu\over{16\pi^2}}\left(-3g_2^2-g_1^2+h_\tau^2+
    3h_b^2+3h_t^2\right)
\end{eqnarray}
and the phase of $\mu$ does not run. As it will be important for our
discussion  below, we also give the RGE for the bilinear mass term $B$
\begin{eqnarray}
{dB \over dt} = {1\over{8\pi^2}}\left(
    -3g_2^2\, M_2-g_1^2\, M_1 + h_\tau^2 A_\tau + 3 h_b^2 A_b + 3 h_t^2 A_t\right)
\label{brge}
\end{eqnarray}
>From Eq. (\ref{brge}), we see that the phase of $A$ will induce a phase
in $B$ (and hence in $\mu$, when we rotate the Higgs fields to ensure
their vacuum expectation values are real).  For this reason, it was argued \cite{gw}
that a tight constraint on $\thm$, required either a similarly tight bound on
$\tha$ or a fine-tuning of the GUT value of $\theta_B$.  

To be more explicit, the Higgs superfield $H_2$ (which gives mass
to up-type fermions) can be phase rotated in such a way so as to insure real
expectation values for the Higgs scalars. The rotation changes the
phase of $H_2$ by an amount $-(\thm + \theta_B)$. Not only is the phase of
$\mu$ now fixed at
$\thm = -\theta_B$, but also the initial phase of $\mu$ is physically
irrelevant as it is canceled by the rotation. As such, one might worry, that
the running of the phase of
$B$, may induce a large phase for $\mu$ when $\tha$ is large, and can be
excluded by the limits on the electron and neutron edms \cite{gw}.  However,
since $\thm =-\thb$, the value of $\thm$ at the weak scale will depend on both $\thb$ and
$\tha$ at the GUT scale. Thus, any value of $\thm$ may be obtained with a
suitable choice of $B$ at the GUT scale for any $\tha$.  Indeed, for large
$\tha$, $\thb$ at the GUT scale is sometimes required to be comparable to 
$\thm$, is sometimes considerably larger, but 
is rarely required to be set
to a precision of  better than 20\% to obtain an acceptable value of $\thm$.
We will return to this point below with some explicit results.

Previously \cite{fkos}, it has been shown that the presence of new
$\cp$-violating phases may have a significant effect on the relic
density of bino-type neutralinos.  The dominant channel for bino
annihilation is into fermion anti-fermion pairs.  However, this
process exhibits p-wave suppression, so that the zero temperature
piece of the thermally averaged annihilation cross-section (which is
relevant for the annihilation of cold binos) is suppressed by a factor
of the final state fermion mass$^2$.  This significantly reduces the
annihilation rate and increases the neutralino relic density.  Mixing
between left and right sfermions lifts this suppression to some extent
by allowing an s-wave contribution to the annihilation cross-section
which is proportional to the bino mass$^2$, but the presence of
complex phases in the off-diagonal components of the sfermion mass
matrices dramatically enhances this effect. 

To fix our notation, we take the general form
of the sfermion mass$^2$ matrix to be \cite{er}
\begin{equation}
\pmatrix{ M_L^2 + m_f^2 + \cos 2\beta (T_{3f} - Q_f\sin ^2 \theta_W) M_Z^2 &
-m_f\,\overline{m}_{\ss f} e^{i \gamma_f}
\cr
\noalign{\medskip} -m_f\,\overline{m}_{\ss f} e^{-i \gamma_f} & M_R^2 + m_f^2 +
\cos 2\beta Q_f\sin ^2
\theta_W M_Z^2
\cr }~
\end{equation}
where $M_{L(R)}$ are the soft supersymmetry breaking sfermion mass
which we have assumed are generation independent and generation
diagonal and hence real.  Due to our choice of phases, there is a
non-trivial phase associated with the off-diagonal entries, which we
denote by $\mf(\overline{m}_{\ss f} e^{i \gamma_f})$, of the sfermion
mass$^2$ matrix, and
\begin{equation}
  \label{mfbar}
  \overline{m}_{\ss f} e^{i \gamma_f} = R_f \mu + A_f^* = R_f
  |\mu|e^{i \theta_\mu} + |A_f|e^{-i \theta_{A_{\ss f}}},
\end{equation}
where $m_f$ is the mass of the fermion $f$ and $R_f =
\cot\beta\:(\tan\beta)$ for weak isospin +1/2 (-1/2) fermions.  We
also define the sfermion mixing angle $\theta_f$ by the unitary matrix
$U$ which diagonalizes the sfermion mass$^2$ matrix,
\begin{equation}
U = \pmatrix{ \cos\theta_{\!f} & \sin\theta_{\!f}\, e^{i\gamma_f} \cr
  \noalign{\medskip}
              -\sin\theta_{\!f} \,e^{-i\gamma_f} & \cos\theta_{\!f}\cr }.
\end{equation}

The effect of $\cp$-violating phases on the annihilation cross section is
manifest through both the sfermion mixing angle and a combination of the
$\cp$-violating phases 
\beq
\sin^2(2\theta_f)\sin^2\gamma_f
\eeq
For the sfermion partners of the light fermions,
\begin{equation}
  \label{sin2thf}
  \sin^2(2\theta_f)\approx {m_f^2\mfb^2\over
                 (M_L^2 - M_R^2 + 2\,Q_f \cos 2\beta\sin^2\!\theta_W M_Z^2)^2}
\end{equation}
Therefore, while in the more general MSSM where there is more freedom to choose
$M_L$ and $M_R$, the effect of the phases on the relic density can be
quite significant, in mSUGRA, the effect is substantially reduced since
$\sin^2(2\theta_f)$ is small for the lighter of generations. (For an LSP with a
mass less than $m_t$, the effect of the phases in the stop mass matrices is
unimportant.)

\begin{figure}
 \begin{center}
\epsfig{file=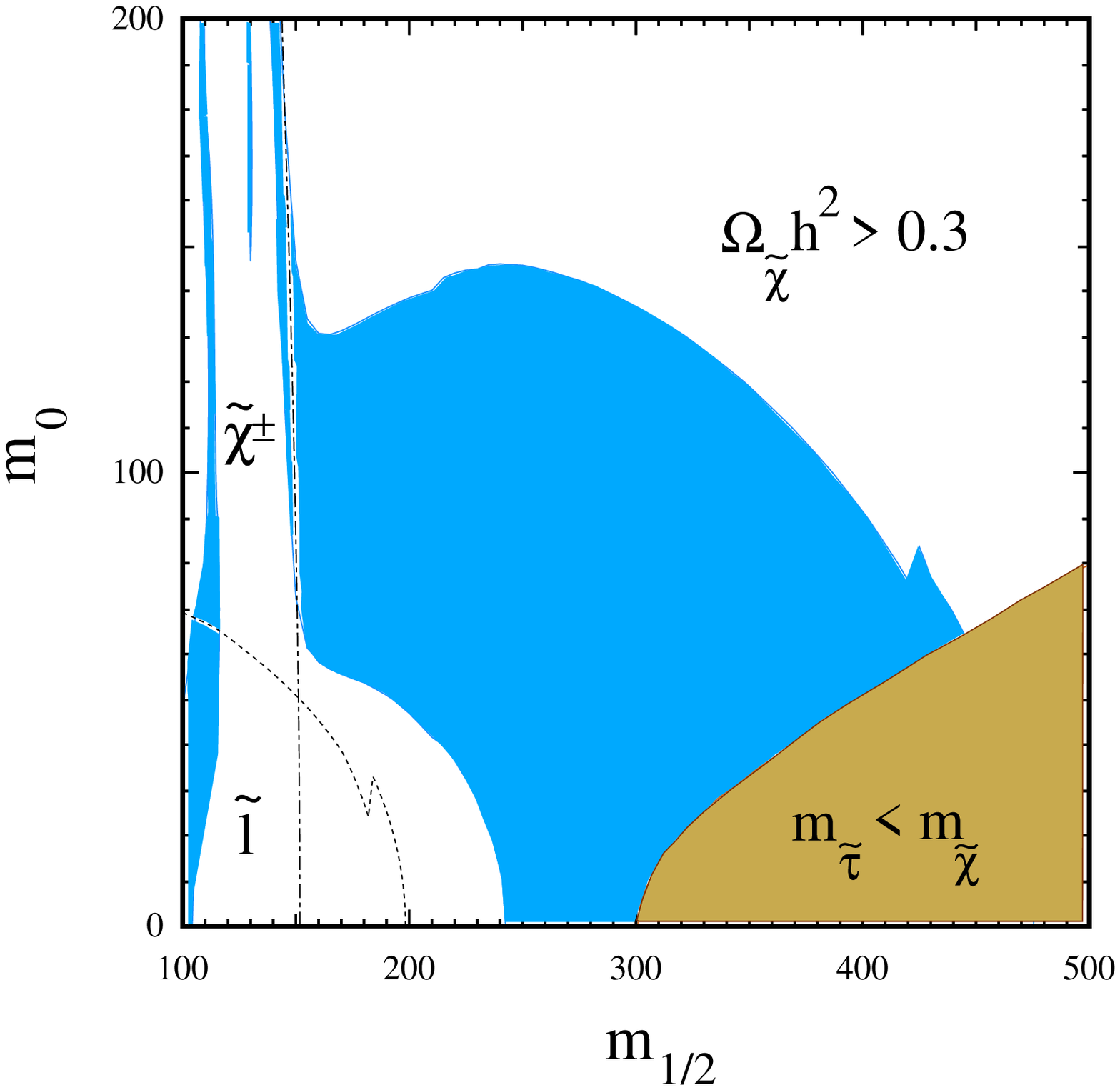,height=7in} 
\vspace{-0.5in}
\caption{\label{fig:rd2c}Contours of constant $\ohsq=0.1$ and $0.3$,
as a function of $m_0$ and $M$, for \mbox{$A_0=0 \gev$} and $\tan\beta=2$.  
The dotted line represents the current LEP2 slepton exclusion contour, and 
the dot-dashed line corresponds to a chargino mass of $91\gev$.  The shaded
region at bottom right yields a stau as the LSP.}
\ \end{center}
\end{figure}

In addition, because of fixed relationship between the many mass
parameters in mSUGRA, the cosmological relic density is able to
furnish a strong constraint on the remaining SUSY parameter space.
For example, in Figure 1, we show the $m_0$-$m_{1/2}$ parameter plane
for $\tan \beta = 2$, the regions for which the relic density takes
the values $0.1 \le \Omega h^2 \le 0.3$.  The upper bound comes from
the requirements that the age of the universe $t_U>12\gyr$ and that
$\Omega\le 1$.  The choppy region at lower values of $m_{1/2}$ shows
the effects of the Higgs and Z poles on the annihilation
cross-section.  Also shown on the figure are the current LEP2 slepton
mass bound\cite{alephslps}, which is about $84\gev$ for large $m_{\tilde l_R}-\mchi$,
and a chargino mass contour of 91 GeV, which approximates the LEP2
bound on the $m_{1/2} - m_0$ plane except near the intersection of
these two contours (see \cite{efos2} for detail).  The hatched regions
in Figure~1 is ruled out because it leads to a stau as the LSP.  As
one can see, the requirement that $\Omega h^2 \le 0.3$ gives us a
constraint on both $m_0$ and $m_{1/2}$: $m_0 \la 150$ GeV, and
$m_{1/2} \la 450$ GeV.  The shape of the allowed region is insensitive
to $\tan\beta$ for small to moderate $\tan\beta$, and these bounds on
$m_0$ and $\m12$ do not vary up to $\tan\beta\sim 8$.  At
$\tan\beta=10$, the bound on $m_0$ is relaxed to $\sim 170\gev$, and
between $\tan\beta\sim 15$ and 20, s-channel pseudo-scalar annihilation
becomes important, and the bounds on both $m_0$ and $\m12$ are
significantly relaxed.  However, the EDMs are typically very large at
these large values of $\tan\beta$ and yield correspondingly tiny upper
bounds on $\thm$.  The cosmological region is also insensitive to
$A_0$, which we take equal to 0 in this plot; the region with
$m_{\tilde\tau}<\mchi$ will vary somewhat with $A_0$ at large
$\tan\beta$, but this has only a small effect on the upper bound on
$\m12$.  In  Fig.~\ref{fig:rd2c} we've taken $\thm=0$, but very similar bounds 
on $m_0$ and $\m12$ apply
for other values of $\thm$.  In the shaded regions, the lightest neutralino is mostly
bino\cite{susygut}, and for large values of the unified gaugino mass
$\m12$, $\chi_1^0$ is almost pure bino.  The cosmological constraints
from this figure will serve as a basis for our constraints on the
$\cp$-violating phases.

The electric dipole moments of the
neutron and the electron has contributions coming from
neutralino, chargino, and gluino exchange to
the quark electric dipole moment. These were calculated in detail in
\cite{ko} and more recently in \cite{in2,in1} correcting some sign errors.
As we will see, this sign change plays an important role in determining
cancellations among the different contributions to the edms.
  The necessary
$C\!P$ violation in these contributions comes from either
$\gf$ in the sfermion mass matrices or $\theta_\mu$ in the neutralino and
chargino mass matrices.  Full expressions for the chargino, neutralino and
gluino exchange contributions are found in \cite{ko,in2}. The neutron edm
has additional contributions to its edm coming from the quark chromoelectric
operator, and a purely gluonic operator.  For all SUSY mass scales of O(100)
GeV, it was shown in \cite{aln} that these latter two contributions to the
quark electric dipole moment are small, coming in with the ratio
$O_\gamma:O_{qc}:O_G = 21:4.5:1$ and hence they were neglected in our previous
analysis.  However as has been argued by Ibrahim and Nath \cite{in2,in1},
these ratios are not general over the interesting supersymmetric parameter
space, and in fact, the latter two contributions can in some instances
even dominate. 

The contributions to the quark electric dipole moments from the
individual gaugino exchange diagrams are proportional to $m_i/m_{\widetilde
q}^2$ where $m_i$ is the mass of the neutralino, chargino or gluino
exchanged.  The edm falls as
$m_{1/2}$ is increased, because the squark masses$^2$ receive large
contributions proportional to
$m_{1/2}^2$ during their RGE evolution from $M_X$ to $M_Z$.  Roughly,
\begin{equation}
  \label{msf}
  m_{\widetilde q}^2 \approx m_0^2 + 6 m_{1/2}^2 + O(M_Z^2).
\end{equation}
The chromoelectric contribution has a similar dependence on SUSY masses.
The gluonic contribution goes as $1/m_{\tilde g}^3$ but still falls as
$m_{1/2}$ is increased if gaugino unification is assumed. Thus even for large
values of the
$C\!P$ violating phases, one can always turn off the quark electric dipole
moment contributions to the neutron EDM (and similarly turn off the electron EDM)
by making $\m12$ sufficiently
large\cite{ko}; however one must still satisfy the cosmological bounds discussed
above.  Experimental bounds are $|d_n| < 1.1
\times10^{-25}e\:{\rm cm}$ \cite{nexp} for the neutron electric dipole
moment and $|d_e| < 4\times10^{-27}e\:{\rm cm}$ \cite{eexp}
for the electron EDM.  Note also that the squark masses $m_{\widetilde q}^2$
are only
weakly dependent on $m_0$ in the cosmologically allowed regions of
Figures~1 and 2, particularly in the regions of large $\m12$, 
 and so the quark EDM's will also be independent of $m_0$.

\begin{figure}
\vspace*{-2.3in}
\begin{minipage}{6.0cm}
\hspace*{-1in}
\epsfig{file=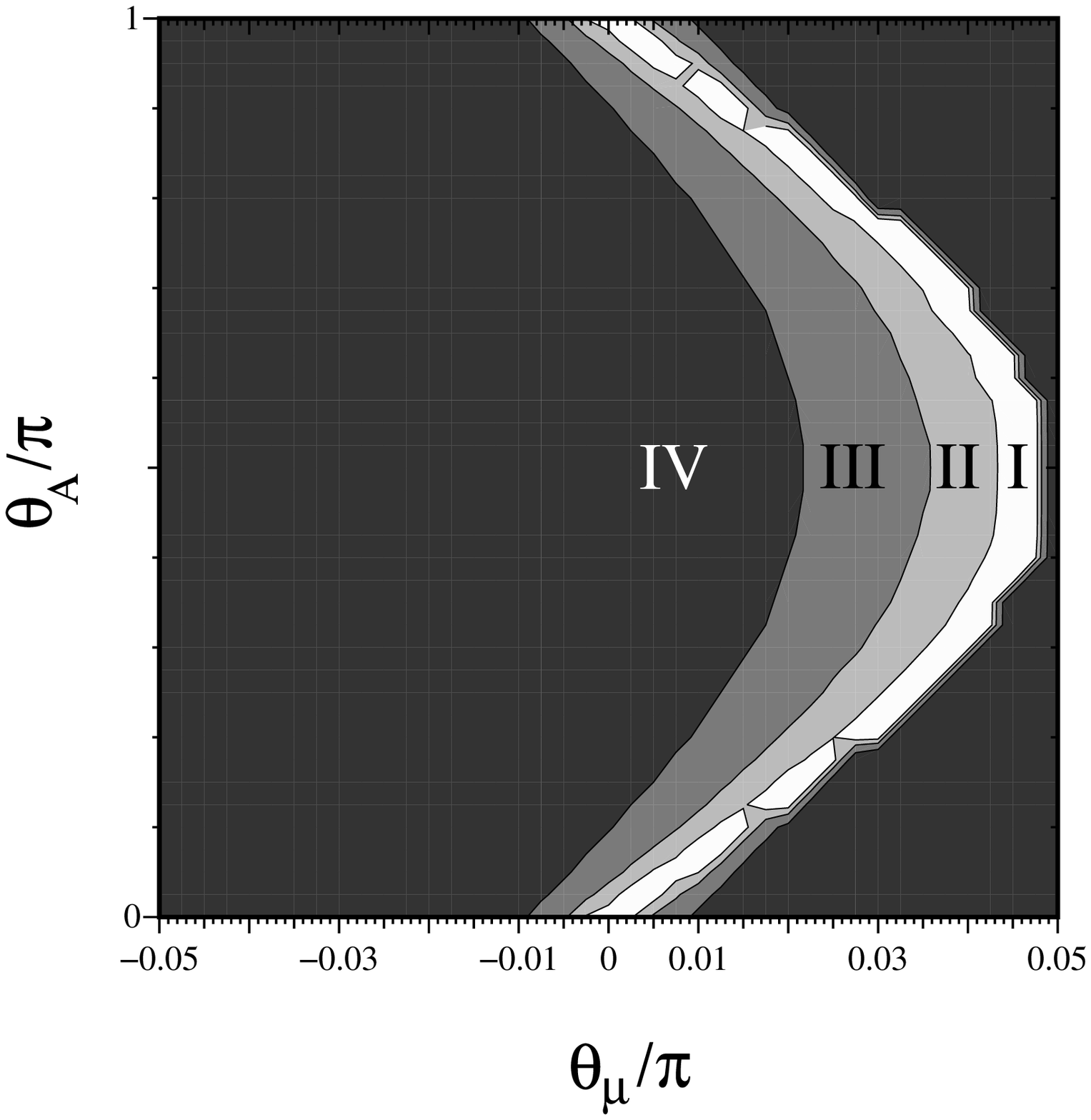,height=6in} 
\end{minipage}
\hspace*{0.3in}
\begin{minipage}{6.0cm}
\epsfig{file=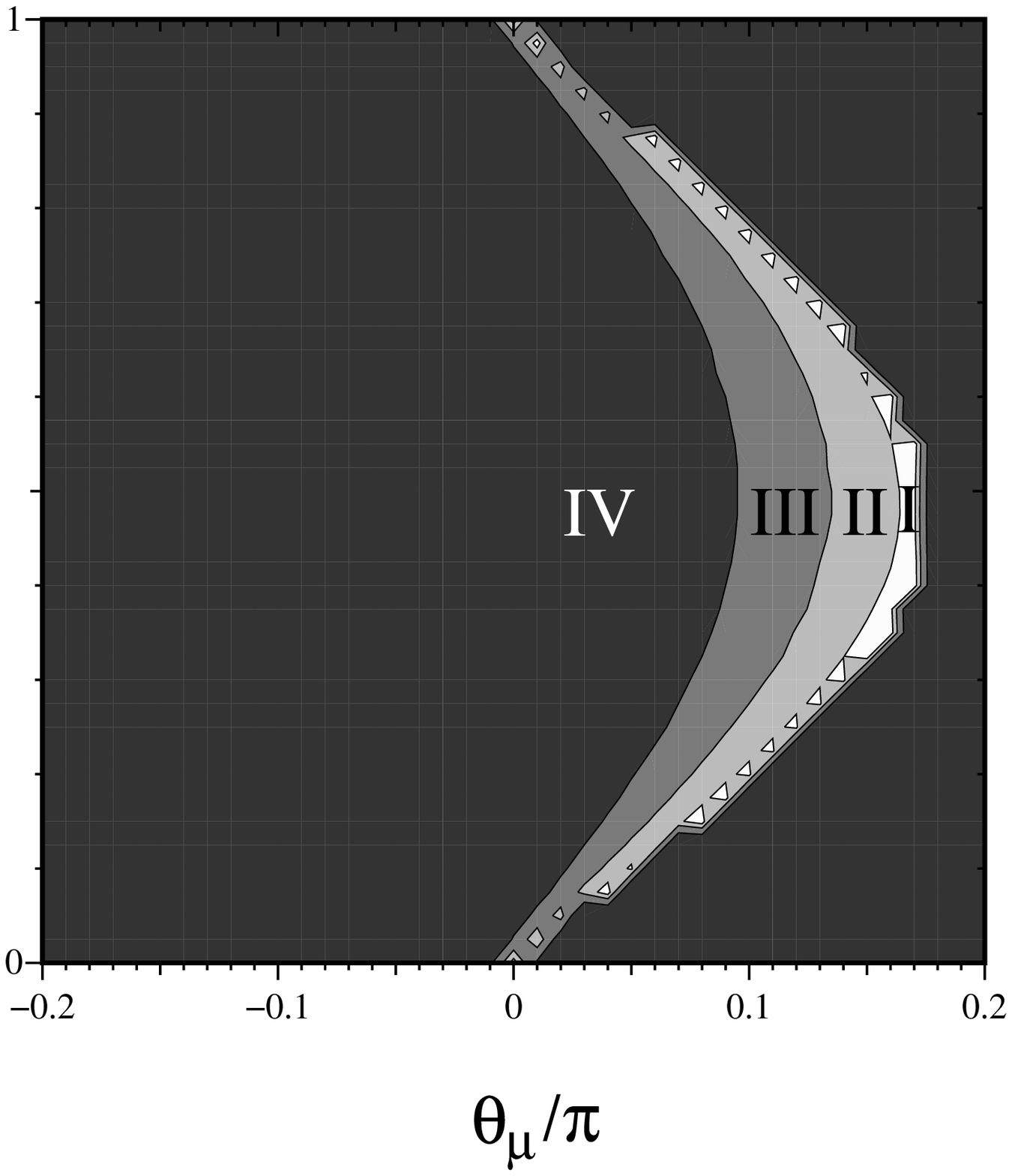,height=6in} 
\end{minipage}\hfill
\vspace{-2.0in}
\begin{minipage}{6.0cm}
\hspace*{-1in}
\epsfig{file=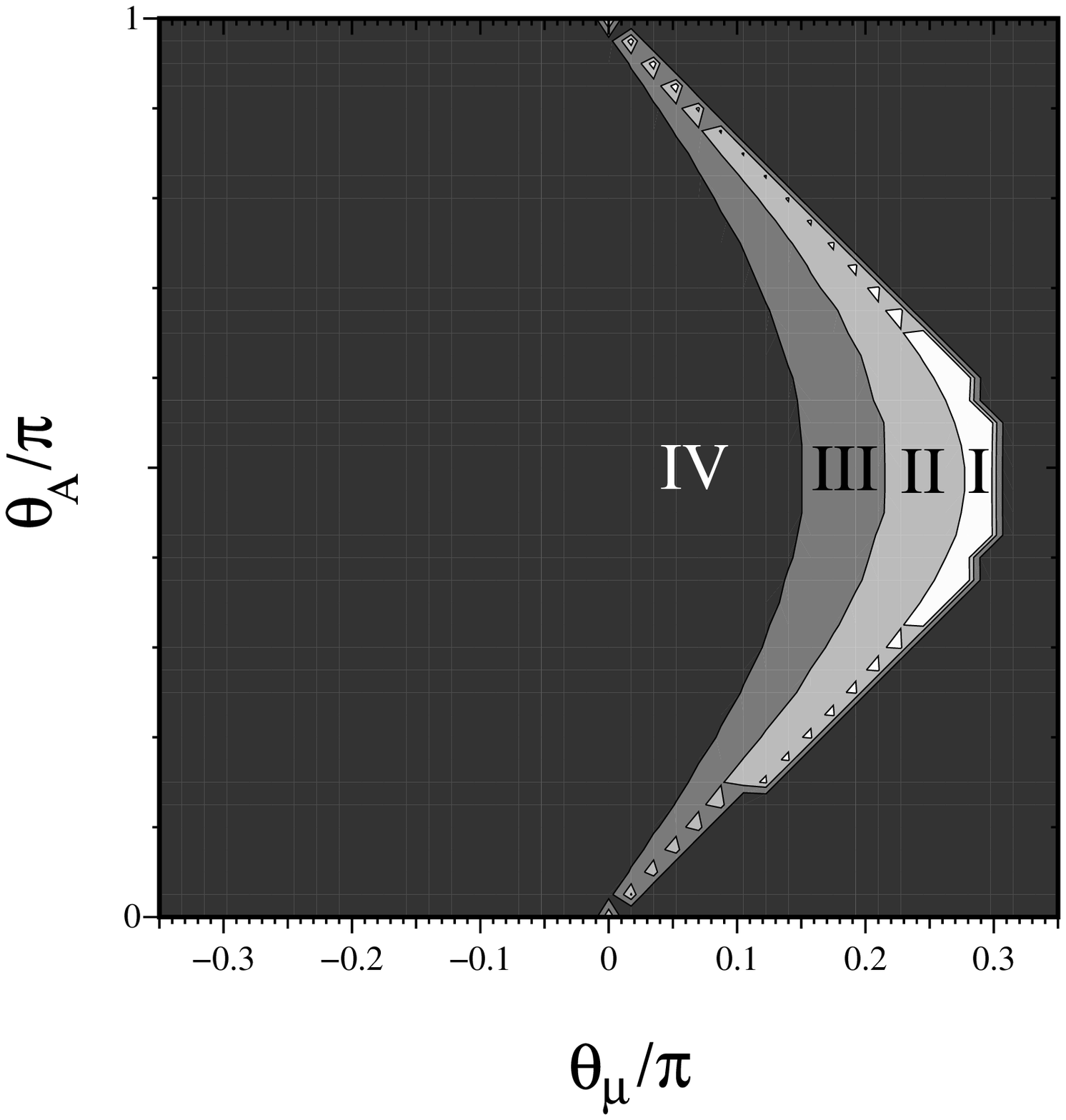,height=6in} 
\end{minipage}
\hspace*{0.3in}
\begin{minipage}{6.0cm}
\epsfig{file=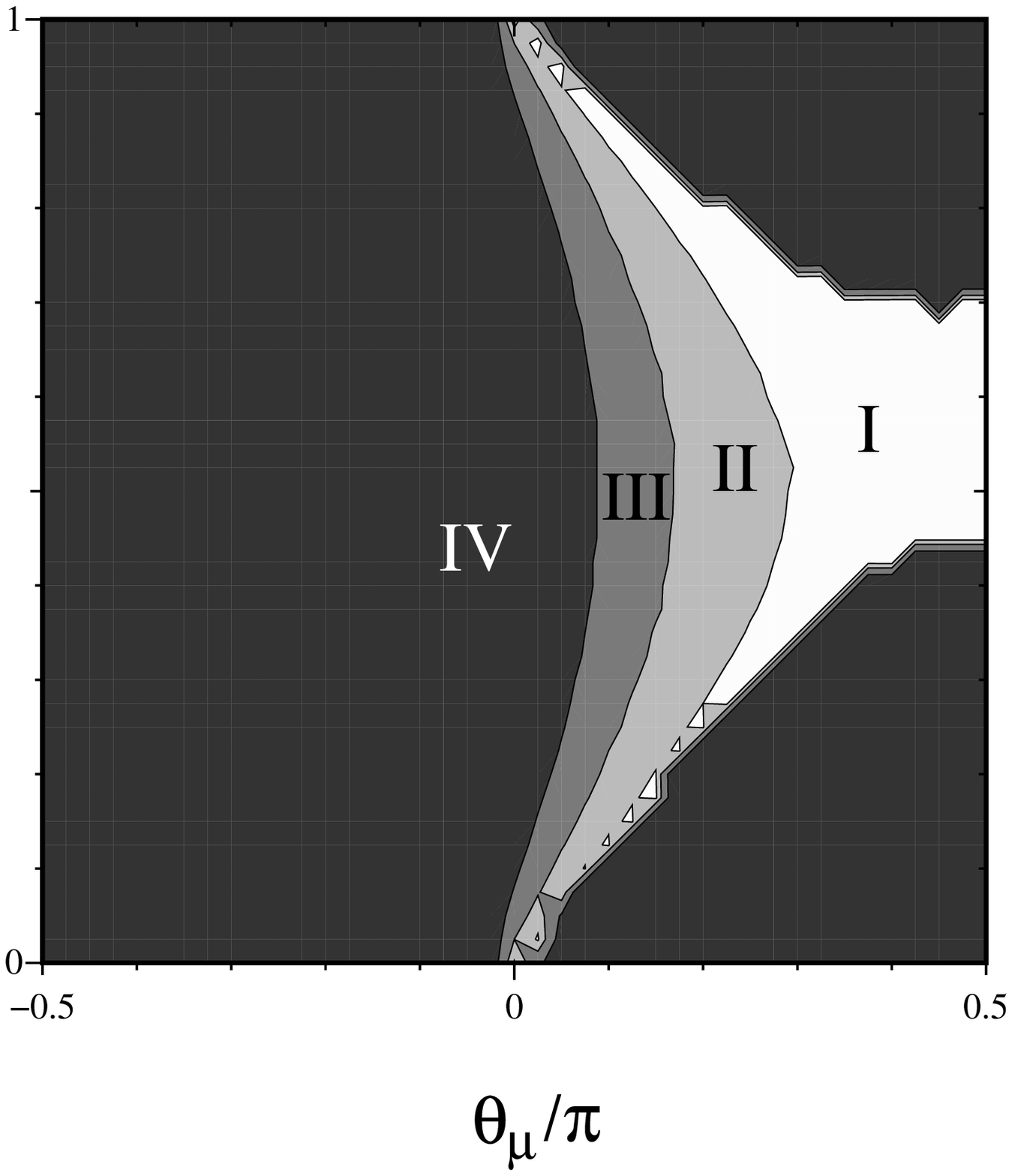,height=6in} 
\end{minipage}\hfill
\vspace{-0.5in}
\caption{\label{fig:eedm}Contours of $\m12^{\rm min}$, the minimum $\m12$ required
to bring the electron EDM below experimental bounds, for $\tan\beta=2, m_0=100\gev$,
and a)$A_0=300\gev$, b)$A_0=1000\gev$, c)$A_0=1500\gev$ and d) as in c)
but for the neutron edm.  The central light zone labeled ``I'' has 
$\m12^{\rm min}<200\gev$, while the zones labeled
``II'', ``III'', and ``IV'' correspond to \hbox{$200\gev<\m12^{\rm min}<300\gev$},
$300\gev<\m12^{\rm min}<450\gev$, and $\m12^{\rm min}>450\gev$, 
respectively.  Zone IV is therefore cosmologically
excluded.}
\end{figure}

We find that, taken independently, 
the current experimental bounds on the electron EDM are generally
more restrictive than the current bounds on the neutron EDM.    We compute
the electron EDM in mSUGRA as a function of  $\thm,
\tha,$ and $m_{1/2}$ for fixed $A_0, m_0, $ and $\tan\beta$.  We find that both the 
neutralino and chargino exchange diagrams can contribute significantly to the electron
EDM, and in fact significant cancellations between the two contributions allow 
for much larger values of $\theta_\mu$ than would be permitted in the absence of
cancellations.   In Fig.~\ref{fig:eedm}, we display as a function of $\thm$ and 
$\tha$ the minimum value of $\m12$ required to bring the electron EDM below the
experimental bounds for $\tan \beta = 2$, taking $m_0=100\gev$.  
We exclude points which violate the current LEP2 chargino and slepton \cite{alephslps}
 mass bounds. 
All the contour plots we present
are computed on a 40x40 grid in $\thm$ and $\tha$, and features smaller than
the grid size are not significant.  Due to cancellations, the electron
EDM does not decrease monotonically with $\m12$ in this region of parameter
space, but there is still a minimum value of $\m12$ which is permitted. 
The cancellations also enhance the dependence of $\m12^{\rm min}$ on
$A_0$.  Since the chargino contribution is proportional to $\sin\gf$,
the value of $\thm$ at which these delicate cancellations occur is
sensitive to $A_0$, and larger values of $A_0$ cause the cancellations
to occur at larger values of $\thm$.   This is demonstrated in
Fig.~\ref{fig:eedm}(a-c), where contours of $\m12^{\rm min}=200, 300$, 
and $450\gev$ are displayed for $A_0= 300, 1000$, and $1500\gev$ respectively. 
In the zone labeled ``I'',  $\m12^{\rm min}<200\gev$, while the zones labeled
``II'', ``III'', and ``IV'' correspond to $200\gev<\m12^{\rm min}<300\gev$,
$300\gev<\m12^{\rm min}<450\gev$, and $\m12^{\rm min}>450\gev$, 
respectively.
For $A_0$=0, the contours are of course straight vertical, and the 
$\m12^{\rm min}=450\gev$ contours lie at $\thm/\pi=\pm\, 0.009$.

The extent of the bowing of the $\m12^{\rm min}$ contours in
Fig.~\ref{fig:eedm} increases with the the fineness of the
cancellation between the neutralino and chargino contributions, and at
the larger values of $\thm$, these cancellations can be very severe,
at the level of one part in 20 for $\thm/\pi$ near $0.05$ in
Fig.~\ref{fig:eedm}a, to the level of one part in a hundred or so in
the most extreme cases in Fig.~\ref{fig:eedm}c.  Since the two
contributions scale very similarly with $\m12$, the cancellations
typically occur over a broad range in $\m12$.  In the extreme cases
the latter is no longer true, as even a tiny relative shift in the
magnitudes of the two contributions is sufficient to spoil the
delicate cancellations necessary to yield a sufficiently small
electron EDM at large $\thm$ and $\tha$.  Numerically, in
Fig.~\ref{fig:eedm}a, the allowed regions are greater than $90\gev$ 
wide in $\m12$ essentially everywhere in the figure, including at large $\thm$
where $\m12^{\rm min}$ is less than $200\gev$.  In
Fig.~\ref{fig:eedm}b, the allowed regions in 
zones I and II are between 20 and $40\gev$ wide in $\m12$
for essentially all $\thm/\pi>0.14$.  And in Fig.~\ref{fig:eedm}c,
 the allowed regions in 
zones I and II are between 10 and $20\gev$ wide for
all $\thm/\pi >0.24$.

Comparing with Fig.~1, we see that regions for which $\m12^{\rm min}$ is
greater than about
$450\gev$ are cosmologically excluded.  For Fig.~\ref{fig:eedm}a, for example,
this restricts
$|\thm|/\pi< 0.05$ for $A_0 = 300 \gev$ (there is an identical region with
$\tha\rightarrow -\tha, \thm\rightarrow -\thm$), while for $A_0 = 1500 \gev$,
$\thm/\pi$ can take values as large as $0.3$.  These large values of $\thm$ are
significantly different from the results found earlier \cite{fko1}.  The differences are
partly due to the sign correction in the neutralino contribution\cite{in1,in2} to the electron
EDM, which
shifts the region of cancellation between the neutralino and chargino contributions,
and partly due to the fact that here we consider larger values of $A_0$.
Of course for larger values of 
$\thm$, the cancellations become finer and finer, and the range of $\m12$ over which 
the electric EDM bounds are satisfied become smaller and smaller.  
The neutron EDM bound will then play a r\^ole in
excluding some  of these small regions, and we will see that the regions which
still remain at large $A_0$ and $\thm$ after imposing both the neutron and
electron EDM bounds have a very small extent in $\m12$.  Consequently, we
neglect the tiny regions in $\m12$ which become available at larger $\thm$
when $A_0>1500\gev$.

We can now explicitly check the degree of fine-tuning needed to have
simultaneously a large value for $\tha$ and an acceptable value for
$\thm$. For the case just considered, when $A_0 = 300 \gev$ and $\tha
= 0.5\, \pi$, $0.02 \la \thm/\pi \la 0.05$ implies that $0.03< \thb/\pi
< 0.05$ at the GUT scale. Namely, $\thb$ is comparable in magnitude to $\thm$ and
must be set only to within 40\%.  As we can plainly see, there is no undue
fine-tuning of $\thb$ which is necessary.    For this value of $A_0$, the change
in $B$ as it is evolved from the GUT scale to the electroweak scale is
much smaller in magnitude than $B$ itself, so $\thb$ cannot change
much during its evolution, and $\thb$ consequently must initially lie in a wedge lying 
in the vicinity of $-\thm$.  Similarly, when $A_0= 1500
\gev$, we see from Fig.~\ref{fig:eedm}c that $0.15 \la \thm/\pi \la
0.3$, which requires $0.18 < \thb/\pi < 0.21$.  Since $B$ runs more for
larger $A_0$, there is actually a finer adjustment needed to get the
appropriate $\thm$ in this case, even though the actual range of
$\thm$ is larger, but here too the adjustment needed hardly constitutes a
fine-tuning.  For  $A_0= 1000 \gev$, the corresponding numbers for 
 $0.1 \la \thm/\pi \la 0.17$ are $0.10 < \thb/\pi < 0.14$.

 We have also computed the neutron EDM in mSUGRA as a function of
 $\thm, \tha,$ and $m_{1/2}$ for fixed $A_0, m_0, $ and $\tan\beta$,
 using the na\"{\i}ve quark model (for the effect of other choices for the quark contributions
to the nucleon spin, see \cite{ef}).  In practice we find that the
 dominant contribution to the neutron electric dipole moment in mSUGRA
 comes typically from the chargino exchange contribution to the
 quark EDMs.  However, the gluino exchange contribution to both the
 quark electric dipole and chromoelectric dipole operators can in some
 regions of parameter space yield neutron EDMs contributions of
 comparable size and opposite sign to the chargino exchange
 contribution, and these cancellations permit even larger values for
 $\thm$\cite{in2}, albeit over a quite narrow range in $\m12$.  In
 Fig.~\ref{fig:eedm}d, we display contours of $\m12^{\rm min}$ allowed
 by the neutron EDM constraint alone for $A_0=1500\gev$ and
 $m_0=100\gev$.  In contrast to the electron EDM, $\thm$ is completely
 unconstrained, although the allowed regions are very small, less than
 $10\gev$ wide in $\m12$ for $\thm/\pi>0.3$ and typically less than
 $5\gev$ wide for $\thm/\pi>0.4$.  The constraints from the neutron
EDM are weaker than the constraints from the electron EDM;  however,
since the cancellations in contributions to the electron and neutron EDMs generally
occur for different ranges of $\m12$, it is useful to combine the two constraints.

\begin{figure}
\vspace*{-2.3in}
\begin{minipage}{6.0cm}
\hspace*{-1in}
\epsfig{file=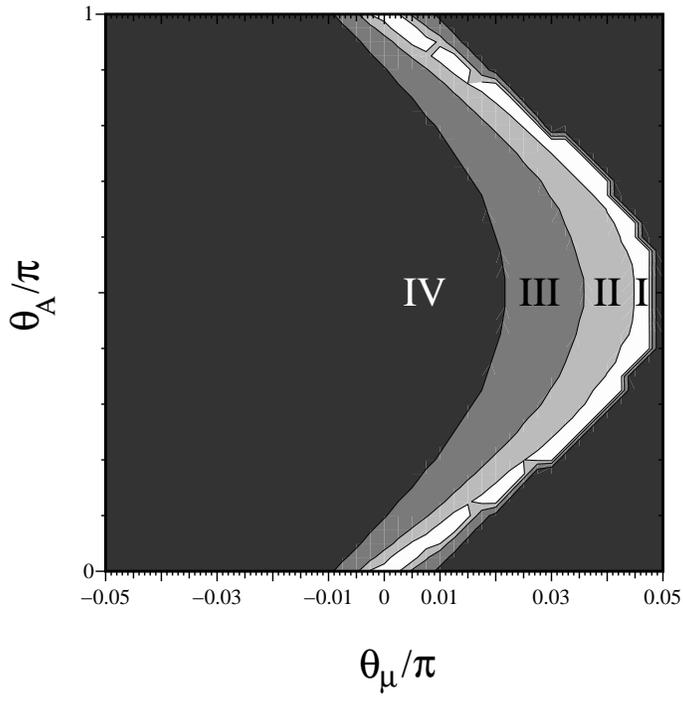,height=6in} 
\end{minipage}
\hspace*{0.3in}
\begin{minipage}{6.0cm}
\epsfig{file=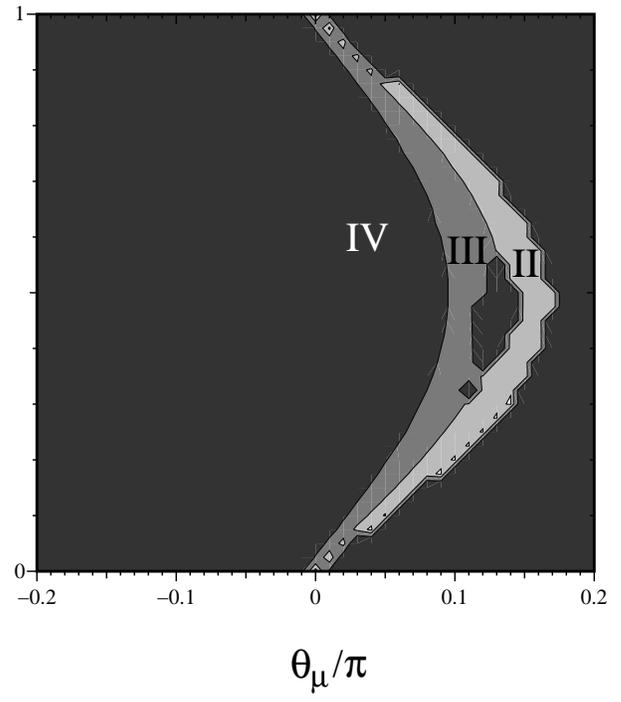,height=6in} 
\end{minipage}\hfill
\vspace{-2.0in}
\begin{minipage}{6.0cm}
\hspace*{-1in}
\epsfig{file=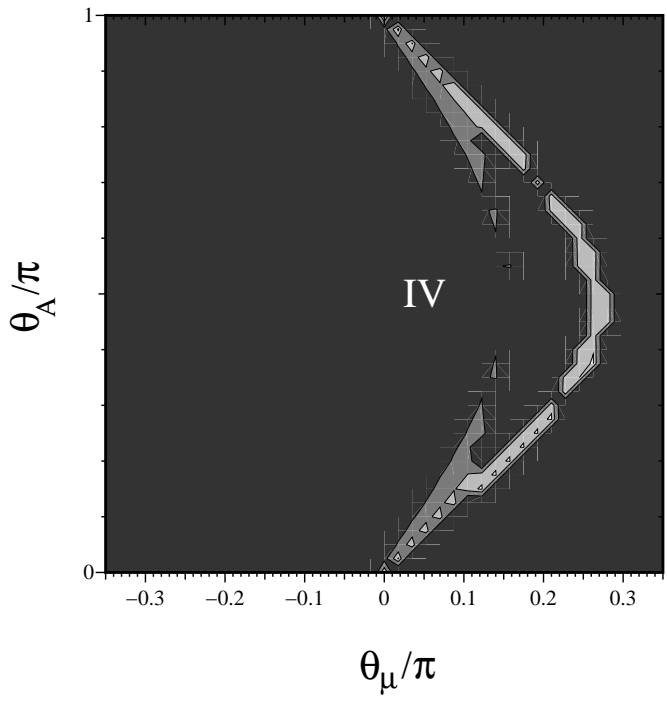,height=6in} 
\end{minipage}\hfill
\caption{\label{fig:bedm}As in Fig.~\protect\ref{fig:eedm}a-\protect\ref{fig:eedm}c, 
but requiring that both the electron 
and  neutron EDM bounds be satisfied.}
\end{figure}

Fig.~\ref{fig:bedm} displays contours of $\m12^{\rm min}$ for the same
parameters as in Fig.~\ref{fig:eedm}, but now requiring that both the
electron and neutron EDM bounds be satisfied.  In
Fig.~\ref{fig:bedm}a, the contours are nearly identical to
Fig.~\ref{fig:eedm}a, as the neutron EDM provides little additional
constraint.  Figs.~\ref{fig:bedm}a and \ref{fig:bedm}b, by contrast,
are noticeably altered.  Essentially all of the regions with
$\m12^{\rm min}< 200\gev$ are removed.  A gap appears in the center of
the allowed region in $\tha$ and $\thm$ , and the gap grows
as $A_0$ is increased.  Further,
the width in $\m12$ of the allowed regions decreases.  In
Fig.~\ref{fig:bedm}c, the width in $\m12$ is between 5 and $15\gev$
for the entire region with $\m12^{\rm min}<300\gev$ and $\thm/\pi >
0.1$.  In Fig.~\ref{fig:bedm}b, the width the allowed region in zone II
varies principally between 10 and $40\gev$.

Thus our general conclusion when taking into account the limits from both the
neutron and electron edm: we find the angle $\tha$ to be unconstrained, that is,
it may take any value between 0 and $\pi$ and still remain consistent with the
neutron and electron edms, as well as with cosmology. However, values of
$\tha = \pi/2$ require  somewhat specific non-zero values of $\thm$ (see
Figs.~\ref{fig:eedm}, \ref{fig:bedm}). As discussed above we do find constraints on $\thm$ which are
quite dependent on the value of $A_0$ ($|A|$ at the GUT scale).  This can be seen by comparing 
Figs.~\ref{fig:eedm}a- \ref{fig:eedm}c and  \ref{fig:bedm}a- \ref{fig:bedm}c,
which show the contours of $m_{1/2}$ in the $\thm - \tha$ plane for $m_0 = 100$ GeV, 
$\tan \beta = 2$.
Recalling from Fig.\ref{fig:rd2c} that $\ohsq<0.3$ requires $m_{1/2}<450\gev$,
one can read off the limits on $\thm$ from the figures.
In \cite{Roc} it was checked that no additional constraint due to the $CP$-violating 
$\epsilon$ 
parameter could be placed on the MSSM phases.

\begin{figure}
 \begin{center}
\vspace*{-0.5in}
\epsfig{file=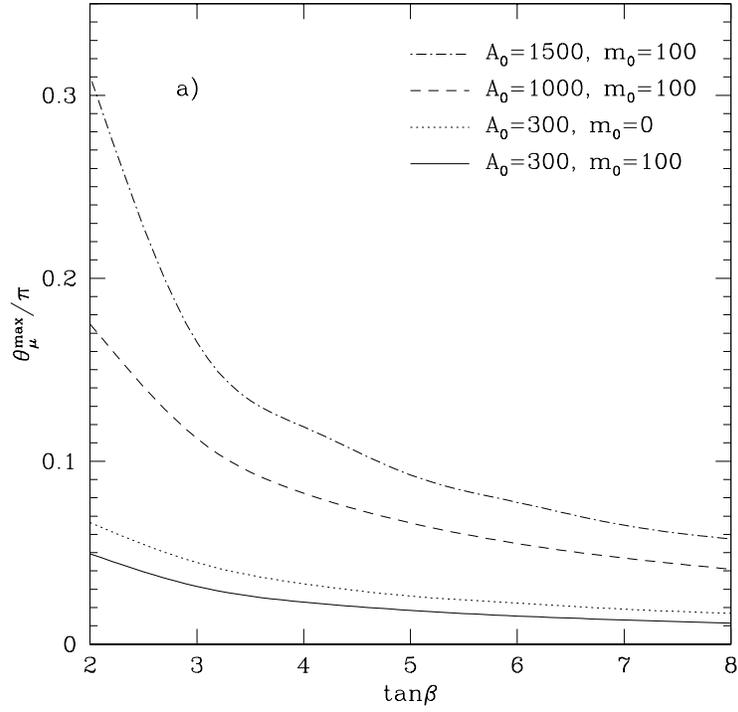,height=4in} 
\epsfig{file=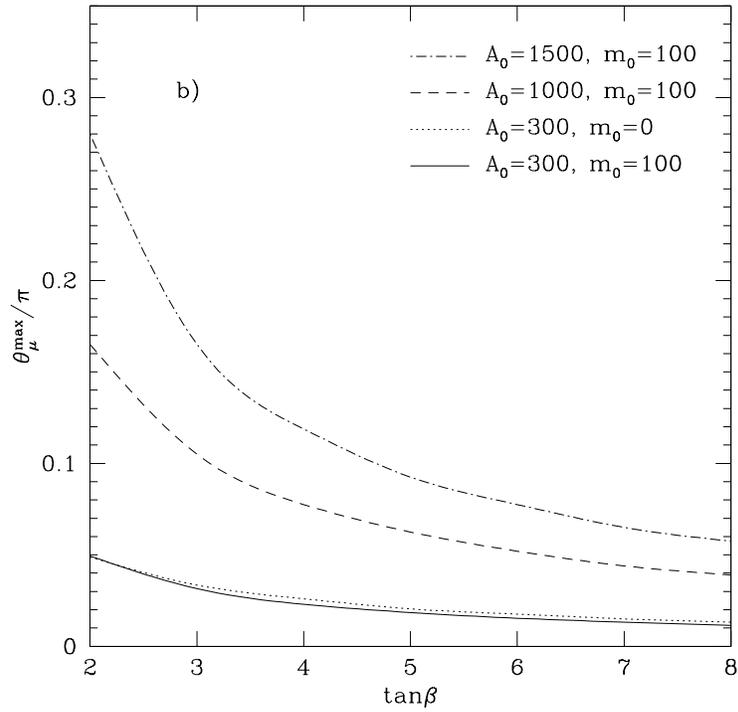,height=4in} 
\caption{\label{fig:thvtb}The maximum values of $\theta_\mu$ allowed by cosmology
and a) the electron EDM and  b) both the electron and neutron EDM's, 
as a function of $\tan\beta$, for several combinations
of $m_0$ and $A_0$.}
\ \end{center}
\end{figure}

The SUSY contributions to the electron and neutron EDMs also depend on
$\tan\beta$.  The contours of constant $\m12^{\rm min}$ in
Figs.~\ref{fig:eedm}, \ref{fig:bedm} keep the the same general shape
as $\tan\beta$ is increased, but as the EDMs do tend to increase with
$\tan\beta$, the scale in $\thm$, and the upper bound on $\thm$,
decreases.  In Fig.~\ref{fig:thvtb}a we display the upper bound on
$\thm$ coming from the electron EDM, as a function of $\tan\beta$ for
four different sets of $A_0$ and $m_0$.  We display the same in
Fig.~\ref{fig:thvtb}b, but here we require that both the neutron and
electron EDM constraints be satisfied.  One sees that varying $m_0$ has only 
a small effect on the $\thm$ bounds in Fig.~\ref{fig:thvtb}a, and a negligible
effect in Fig.~\ref{fig:thvtb}b.

In summary, we have combined cosmological bounds on gaugino masses with
experimental bounds on the neutron and electron electric dipole moments
to constrain the new $\cp$-violating phases in the Constrained Minimal
Supersymmetric Standard Model. We find that the inclusion of the 
chromoelectric and purely gluonic
contributions to the quark electric dipole moment as calculated by Ibrahim
and Nath \cite{in2,in1}, while providing significant corrections to the 
neutron EDM in regions of the parameter space, do not significantly
affect the upper bound on the phases in mSUGRA, from the combination of 
the neutron and electron EDM and relic density constraints. The phase of the
supersymmetric Higgs mixing mass is constrained by $|\thm|/\pi\la 0.3$ for $A_0<1500\gev$ and
$|\thm|/\pi \la 0.05$ for $A_0<300\gev$.
 In addition, 
there is no bound on the phase
$\tha$ of the unified trilinear scalar mass parameter~$A$.

\vskip .3in
\vbox{
\noindent{ {\bf Acknowledgments} } \\
\noindent  The work of K.O. was supported in part by DOE grant
DE--FG02--94ER--40823.  The work of T.F. was supported in part by DOE   
grant DE--FG02--95ER--40896 and in part by the University of Wisconsin  
Research Committee with funds granted by the Wisconsin Alumni Research  
Foundation.}


\newpage

\newpage

\end{document}